\documentclass[aps,epsfig,floatfix,twocolumn,nofootinbib,showpacs]{revtex4}
\usepackage{graphicx}
\usepackage{amsmath}
\usepackage{amsfonts}
\begin{document}

\title{Effect of Disorder Strength on Optimal Paths\\ in Complex
Networks}

\author{Sameet Sreenivasan,$^1$ Tomer Kalisky, $^2$ Lidia
A. Braunstein,$^{1,3}$\\  Sergey V. Buldyrev,$^1$ Shlomo Havlin,$^{1,2}$
and H. Eugene Stanley$^1$}

\affiliation{$^1$Center for Polymer Studies and Department of Physics\\
Boston University, Boston, MA 02215, USA\\ $^2$Minerva Center and
Department of Physics\\ Bar-Ilan University, 52900 Ramat-Gan, Israel\\
$^3$Departamento de F\'{\i}sica, Facultad de Ciencias Exactas y
Naturales\\ Universidad Nacional de Mar del Plata\\ Funes 3350, $7600$
Mar del Plata, Argentina\\ }

\date{7 May 2004}

\begin{abstract}

We study the transition between the strong and weak disorder regimes in
the scaling properties of the average optimal path $\ell_{\rm opt}$ in a
disordered Erd\H{o}s-R\'enyi (ER) random network and scale-free (SF)
network. Each link $i$ is associated with a weight $\tau_i\equiv\exp(a
r_i)$, where $r_i$ is a random number taken from a uniform distribution
between $0$ and $1$ and the parameter $a$ controls the strength of the
disorder. We find that for any finite $a$, there is a crossover network
size $N^*(a)$ at which the transition occurs. For $N \ll N^*(a)$ the
scaling behavior of $\ell_{\rm opt}$ is in the strong disorder regime,
with $\ell_{\rm opt} \sim N^{1/3}$ for ER networks and for SF networks
with $\lambda \ge 4$, and $\ell_{\rm opt} \sim
N^{(\lambda-3)/(\lambda-1)}$ for SF networks with $3 < \lambda < 4$. For
$N \gg N^*(a)$ the scaling behavior is in the weak disorder regime, with
$\ell_{\rm opt}\sim\ln N$ for ER networks and SF networks with $\lambda
> 3$. In order to study the transition we propose a measure which
indicates how close or far the disordered network is from the limit of
strong disorder. We propose a scaling ansatz for this measure and demonstrate
its validity. We proceed to derive the scaling
relation between $N^*(a)$ and $a$. We find that $N^*(a)\sim
a^3$ for ER networks and for SF networks with $\lambda\ge 4$, and
$N^*(a)\sim a^{(\lambda-1)/(\lambda-3)}$ for SF networks with $3 <
\lambda < 4$.

\end{abstract}

\maketitle

\section{Introduction}\label{int}

The subject of complex networks has been widely explored in the past few
years in part due to its broad range of applications to social,
biological and communication systems
\cite{Albert02,Barabasi02,Buchanan02,Watts03,Dorogovtsev03,Pastor}. In a
real world network, whether it be a communication network or transport
network, the time $\tau_i$ taken to traverse a link $i$ may not be the
same for all the links. In other words, there is a ``cost'' or a
``weight'' $\tau_i$ associated with each link, and the larger the weight
on a link, the harder it is to traverse this link. In such a case, the
network is said to be disordered.

Consider two nodes $A$ and $B$ on such a disordered network. In general,
there will be a large number of paths connecting $A$ and $B$. Among
these paths, there is usually a single path for which the sum of the
costs $\sum\tau_i$ along the path is minimum and this path is called the
``optimal path.'' The problem of optimal paths on networks is of import
since the purpose of many real networks is to provide an efficient
traffic route between its nodes.

When most of the links on the path contribute to the sum, the system is
said to be ``weakly disordered'' (WD).  In some cases, however, the cost
of a single link along the path dominates the sum. In this case every
path between two nodes can be characterized by a value equal to the
maximum cost along that path, and the path with the minimal value of the
maximum cost is the optimal path between the two nodes. This limit of
disorder is called the {\it strong disorder\/} (SD) limit (``ultrametric''
limit) \cite{Cieplak} and we refer to the optimal path in this limit as
the {\it min-max path}.

The procedure to implement disorder on a network is as follows
\cite{Cieplak,Porto99,Brauns01,Brauns03}. One assigns to each link $i$
of the network a random number $r_{i}$, uniformly distributed between 0
and 1. The cost associated with link $i$ is then
\begin{equation}
\tau_i\equiv\exp(ar_i),
\label{ee1}
\end{equation}
where $a$ is the parameter which controls the broadness of the
distribution of link costs. The parameter $a$ represents the strength of
disorder. The limit $a\rightarrow\infty$ is the strong disorder limit,
since for this case only one link dominates the cost of the path.

There are distinct scaling relationships between the length of the
average optimal path $\ell_{\rm opt}$ and the network size (number of
nodes) $N$ depending on whether the network is strongly or weakly
disordered \cite{Brauns03}. For strong disorder \cite{Brauns03},
$\ell_{\rm opt} \sim N^{\nu_{\rm opt}}$, where $\nu_{\rm opt} = 1/3$ for
Erd\H{o}s-R\'enyi (ER) random networks \cite{ER59} and for scale-free
(SF) \cite{Albert02} networks with $\lambda > 4$, where $\lambda$ is the
exponent characterizing the power law decay of the degree
distribution. For SF networks with $3 < \lambda < 4$, $\nu_{\rm
opt}=(\lambda-3)/(\lambda-1)$. For weakly disordered ER networks and for
SF networks with $\lambda > 3$, $\ell_{\rm opt} \sim \ln N $. Porto et
al.~\cite{Porto99} considered the optimal path transition from weak to
strong disorder for 2-D and 3-D lattices, and found a crossover in the
scaling properties of the optimal path that depends on the disorder
strength $a$, as well as the lattice size $L$.

Here we show that similar to regular lattices, there exists for any
finite $a$, a crossover network size $N^*(a)$ such that for $N \ll
N^*(a)$, the scaling properties of the optimal path are in the strong
disorder regime while for $N \gg N^*(a)$, the network is in the weak
disorder regime. We evaluate the function $N^*(a)$. The structure of the
paper is as follows. In Section \ref{sc} we derive a scaling approach
for the transition from weak disorder to strong disorder of the optimal
path. In Section \ref{sm} we present simulation results which support
the scaling Ansatz. Finally, in Section \ref{disc} we conclude with an
analytic justification for the scaling of the transition.

\begin{figure}
\includegraphics[width=8.0cm,height=5.0cm,angle=0]{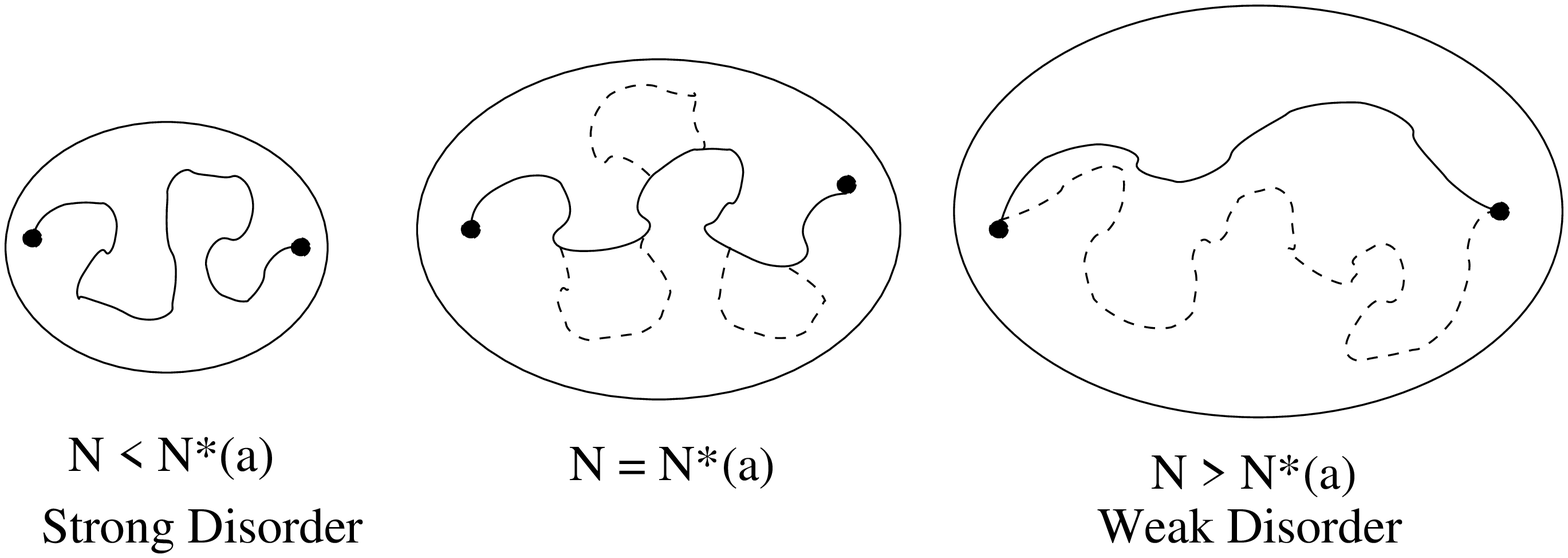}
\caption{Schematic representation of the transition in the topology of
  the optimal path with system size $N$ for a given disorder strength
  $a$. The solid line shows the optimal path at a finite value of $a$
  connecting two nodes indicated by the filled circles. The portion of
  the min-max path that is distinct from the optimal path is indicated
  by the dashed line. (a) For $N \ll N^*(a)$ (i.e. $\ell_{\infty} \ll \ell^*(a)$), the
  optimal path coincides with the min-max path, and we expect the
  statistics of the SD limit. (b) For $N = N^*(a)$ (i.e. $\ell_{\infty} = \ell^*(a)$),
  the optimal path starts deviating from the min-max path. (c) For $N
  \gg N^*(a)$ (i.e. $\ell_{\infty} \gg \ell^*(a)$), the optimal path has almost no
  links in common with the min-max path, and we expect the statistics of
  the WD limit.}
\label{f.1}
\end{figure}
\section{Scaling Approach}\label{sc}

In general, the average optimal path length $\ell_{opt}(a)$ in a
disordered network depends on $a$ as well as on $N$. In the following we
use instead of $N$ the min-max path length $\ell_{\infty}$ which is
related to $N$ as  $\ell_\infty\equiv\ell_{\rm opt}(\infty)\sim
N^{\nu_{\rm opt}}$ and hence $N$ can be expressed as a function of
$\ell_{\infty}$,
\begin{equation}
N\sim\ell_\infty^{1/\nu_{\rm opt}}.
\label{e1}
\end{equation}
Thus, for finite $a$, $\ell_{\rm opt}(a)$ depends on both $a$ and
$\ell_{\infty}$. We expect that there exists a crossover length
$\ell^*(a)$, corresponding to the crossover network size $N^*(a)$, such
that (i) for $\ell_\infty\ll\ell^*(a)$, the scaling properties of
$\ell_{\rm opt}(a)$ are those of the strong disorder regime, and (ii)
for $\ell_\infty\gg\ell^*(a)$, the scaling properties of $\ell_{\rm
opt}(a)$ are those of the weak disorder regime. In Fig.~\ref{f.1}, we
show a schematic representation of the change of the optimal path as the
network size increases.
                                
In order to study the transition from strong to weak disorder, we
introduce a measure which indicates how close or far the disordered
network is from the limit of strong disorder. A natural measure is the
ratio
\begin{equation}
W(a)\equiv\frac{\ell_{\rm opt}(a)}{\ell_\infty}.
\label{equation1}
\end{equation}
Using the scaling relationships between $\ell_{\rm opt}(a)$ and $N$ in
both regimes, and $\ell_{\infty} \sim N^{\nu_{\rm opt}}$ (see Section
\ref{int}), we get
\begin{equation}
\ell_{\rm opt}(a)\sim
\begin{cases}
\ell_\infty\sim N^{\nu_{\rm opt}} & \text{[SD]}\cr \ln\ell_\infty\sim\ln
N           & \text{[WD].}
\end{cases}
\label{equation2}
\end{equation}
From Eq.~(\ref{equation1}) and  Eq.~(\ref{equation2}) it follows,
\begin{equation}
W(a)\sim
\begin{cases}
{\rm const.}                    & \text{[SD]}\cr \ln\ell_\infty/\ell_\infty &
\text{[WD].}
\end{cases}
\label{equation3}
\end{equation}
We propose the following scaling Ansatz for $W(a)$,
\begin{equation}
W(a)=F\left({\ell_\infty\over\ell^*(a)}\right),
\label{equation4}
\end{equation}
where
\begin{equation}
F(u)\sim
\begin{cases}
{\rm const}.  & \text{$u\ll 1$}\cr \ln(u)/u & \text{$u\gg 1$,}
\end{cases}
\label{equation5}
\end{equation}
with
\begin{equation}
\label{e2}
u\equiv{\ell_\infty\over\ell^{*}(a)}.
\end{equation}
The dependence of $\ell^*(a)$ on $a$ can be estimated as follows. In the
strong disorder limit, the costs on the links for any path on the
network typically differ by at least an order of magnitude. This means
that for a min-max path of $\ell$ links (or length $\ell$), if we
arrange the costs of the links in descending order, then two consecutive
costs typically differ at least by an order of magnitude. If $r_1$ and
$r_2$ are the random numbers associated with two such consecutive links,
with $r_1 > r_2$, then the ratio of the costs on the links is
\begin{equation}
\label{e3}
{\tau_1\over\tau_2}=\exp(a\Delta r),
\end{equation}
where $\Delta r\equiv r_1-r_2$. Thus, in the case of strong disorder we
must have $a\Delta r\gg 1$. Consequently the transition to weak disorder
occurs when all the links become equivalent in order of magnitude, i.e.,
when $a\Delta r\sim 1$. The value of $\Delta r$ depends on the length of
the path. If the distribution of random numbers on the min-max path is
uniform, then $\Delta r\sim 1/\ell$ for a min-max path of length
$\ell$. The condition for the transition, $a\Delta r\sim 1$ is satisfied
at the crossover length $\ell^*(a)$ which implies that
\begin{equation}
\label{e4}
\ell^*(a)\sim a.
\end{equation}
Therefore, from Eq.~(\ref{equation4}), $W(a)$ must be a function of
$\ell_\infty/a$.

\section{Simulation Results}\label{sm}

Next we describe the details of our numerical simulations and show that
the results agree with our theoretical predictions. To construct an ER
network of size $N$ with average node degree $\langle k \rangle$, we
start with $\langle k \rangle N/2$ edges and randomly pick a pair of
nodes from the total possible $N(N-1)/2$ pairs to connect with each
edge. The only condition we impose is that there cannot be multiple
edges between two nodes.

In order to generate SF networks, we use the Molloy-Reed algorithm
\cite{Molloy}. Each node is assigned a random integer $k$ taken from a
power-law distribution
\begin{equation}
\label{ee4}
P(k) = \left({k\over k_0}\right)^{-\lambda},
\end{equation}
where $k_0$ is the minimal possible number of links that a node
possesses. Next, we randomly select a node and attempt to connect each
of its $k$ links with randomly selected $k$ nodes that still have free
positions for links. The disorder in the link costs is then implemented
using the procedure described in \cite{Brauns01}.
\begin{figure}
\includegraphics[width=6.0cm,height=8.0cm,angle=270]{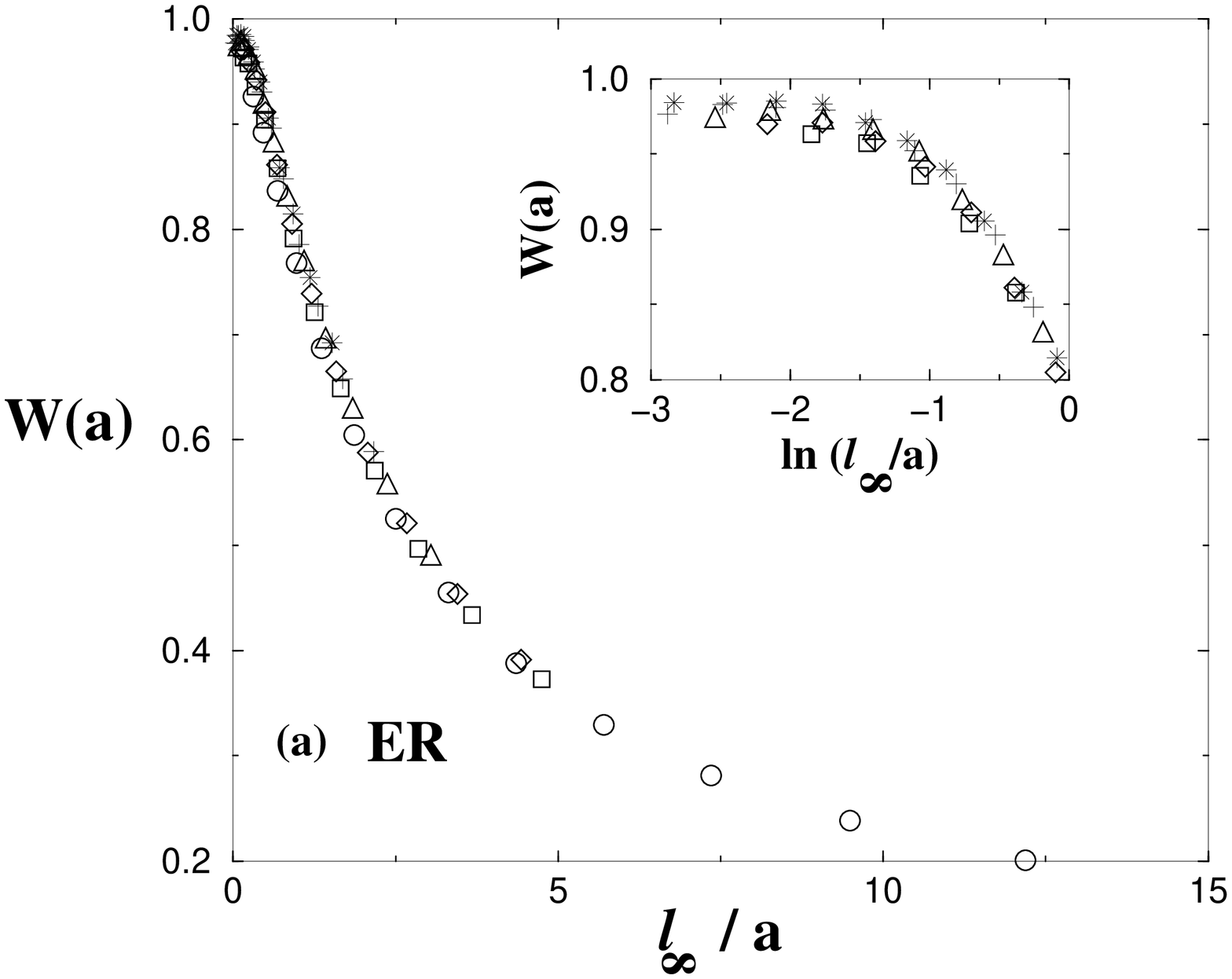}
\includegraphics[width=6.0cm,height=8.0cm,angle=270]{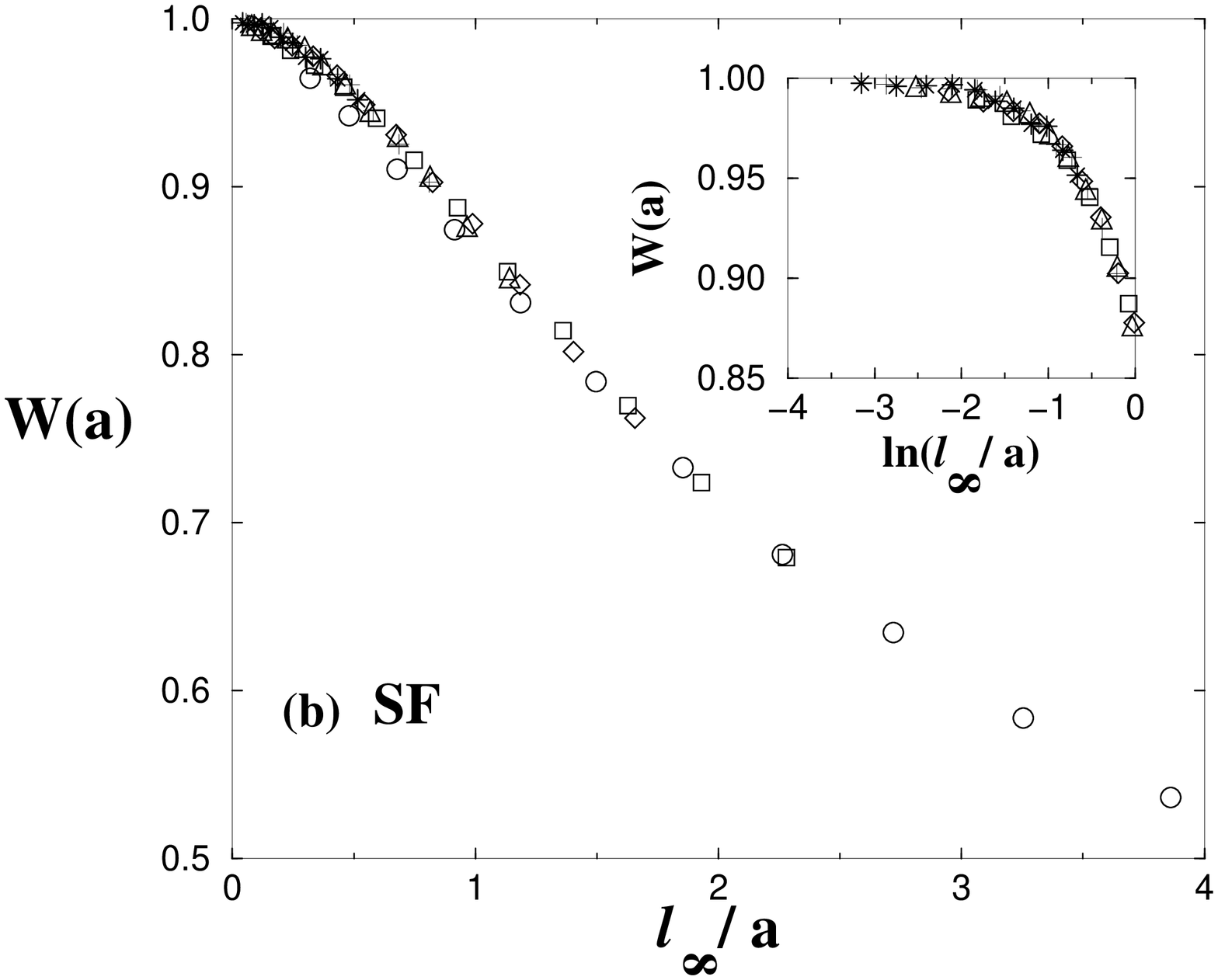}
\caption{Test of Eqs.~(\ref{equation4}) and (\ref{equation5}). (a)
   $W(a)$ plotted as a function of $\ell_{\infty}/a$ for different
   values of $a$ for ER networks with $\langle k \rangle = 4$. The
   different symbols represent different $a$ values: $a = 8$($\circ$),
   $a = 16$($\Box$), $a = 22$($\diamond$), $a = 32$($\bigtriangleup$),
   $a = 45$($+$), and $a = 64$($\ast$). (b) Same for SF networks with
   $\lambda = 3.5$. The symbols correspond to the same values of
   disorder as in (a). The insets show $W(a)$ plotted against
   $\log(l_{\infty}/a)$, and indicate for $\ell_\infty\ll a$, $W(a)$
   approaches a constant in agreement with Eq.~(\ref{equation5}).}
\label{f.2}
\end{figure}
To obtain $\ell_\infty$, we use the algorithm
proposed by Cieplak et al.~\cite{Cieplak}, modified as described in
Ref.~\cite{Brauns203}. With this modification we reach system sizes of
$N=2^{16}=65~536$. In order to obtain the optimal path for a given
realization, we use the Dijkstra algorithm \cite{Porto99}. We calculate
the average optimal path $\ell_{\rm opt}(a)$ by taking the average of
the optimal paths over $10^6$ pairs of nodes.

\begin{figure}
\includegraphics[width=6.0cm,height=8.0cm,angle=270]{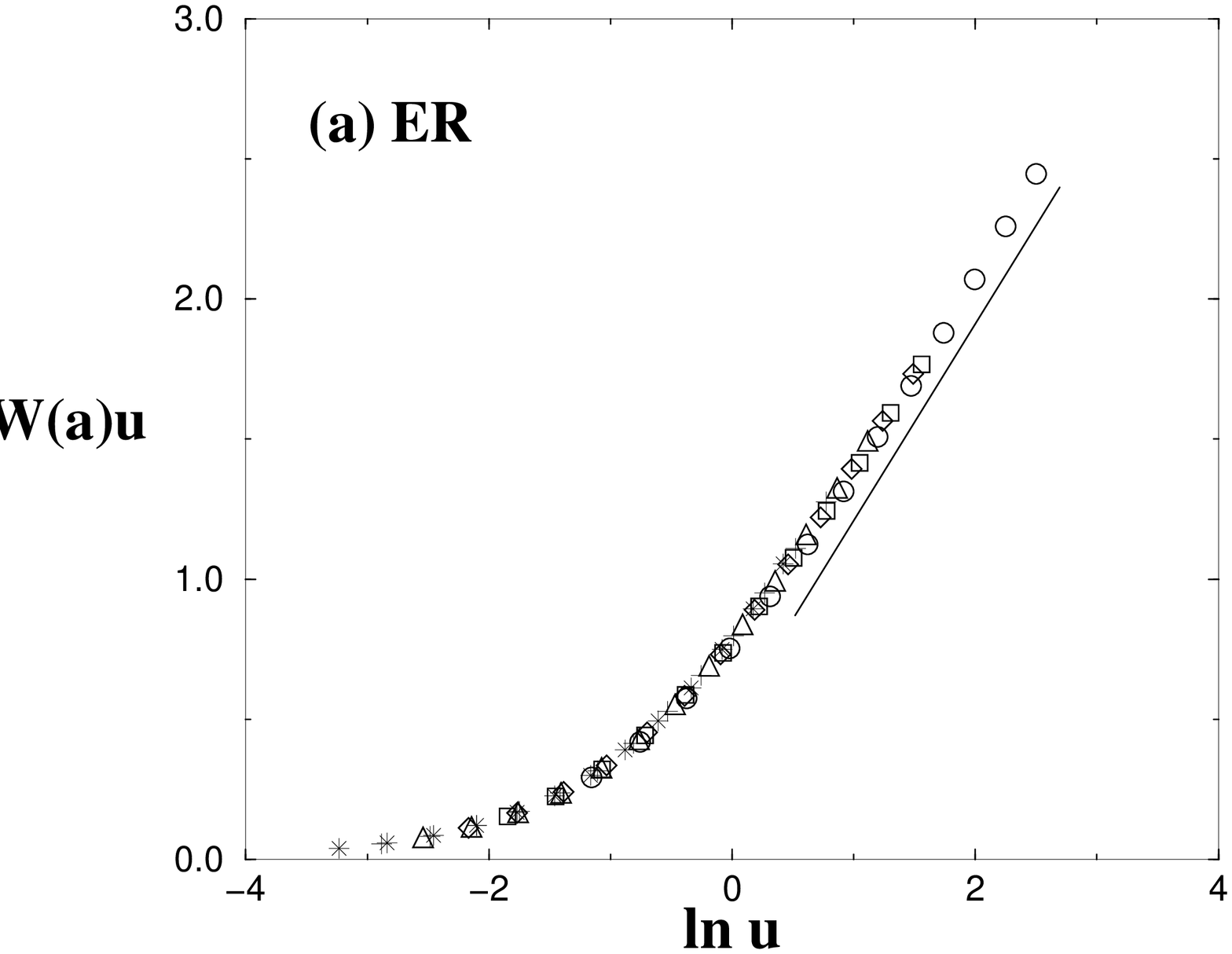}
\includegraphics[width=6.0cm,height=8.0cm,angle=270]{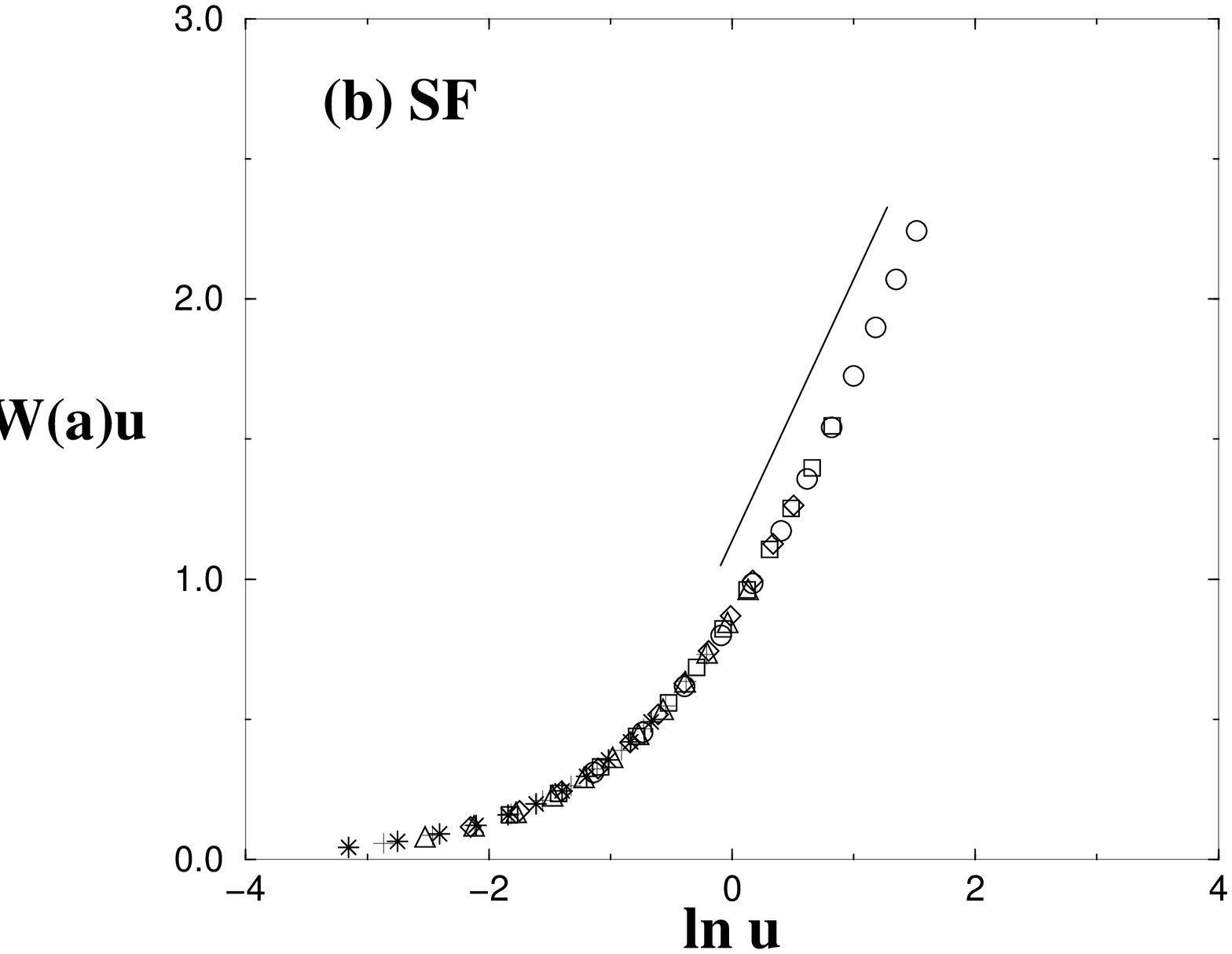}
\caption{(a) Plot of $W(a)u = \ell_{\rm opt}(a)/\ell^*(a) = \ell_{\rm
  opt}(a)/a $ vs $\ln u \equiv \ln (\ell_{\infty}/\ell^*(a))=
  \ln(\ell_{\infty}/a) $ for ER networks with $\langle k \rangle = 4$
  for different values of $a$. (b) Plot of $W(a)u = \ell_{\rm
  opt}(a)/\ell^*(a) = \ell_{\rm opt}(a)/a $ vs $\ln u= \ln
  (\ell_{\infty}/\ell^*(a))= \ln(\ell_{\infty}/a)$ for SF networks with
  $\lambda = 3.5$. The values of $a$ represented by the symbols in (a)
  and (b) are the same as in Fig.~(\ref{f.2}).}
\label{f.3}
\end{figure}

In Fig.~\ref{f.2} we show the ratio $W(a)$ for different values of $a$
plotted against $\ell_\infty/\ell^*(a)\equiv\ell_\infty/a$ for ER networks
with $\langle k \rangle = 4$ and for SF networks with $\lambda=3.5$. The
excellent data collapse is consistent with the scaling relations
Eq.~(\ref{equation4}).
Fig.~\ref{f.3} shows the scaled quantities $W(a)u=\ell_{\rm opt}(a)/\ell^*(a)$ vs. $\ln
u\equiv\ln(\ell_\infty/\ell^*(a))\equiv\ln(\ell_\infty/a)$, for both ER networks
with $\langle k \rangle = 4$ and for SF networks with $\lambda=3.5$. The
curves are linear at large $u\equiv\ell_\infty/\ell^*(a)$, supporting the
validity of the logarithmic term in Eq.~(\ref{equation5}) for large $u$.

\section{Discussion}\label{disc}

We next develop analytic arguments that support Eq.~(\ref{e4}). These arguments will lead to a clearer picture about
the nature of the transition of the optimal path with disorder strength.

\begin{figure}
\includegraphics[width=6.0cm,height=8.0cm,angle=270]{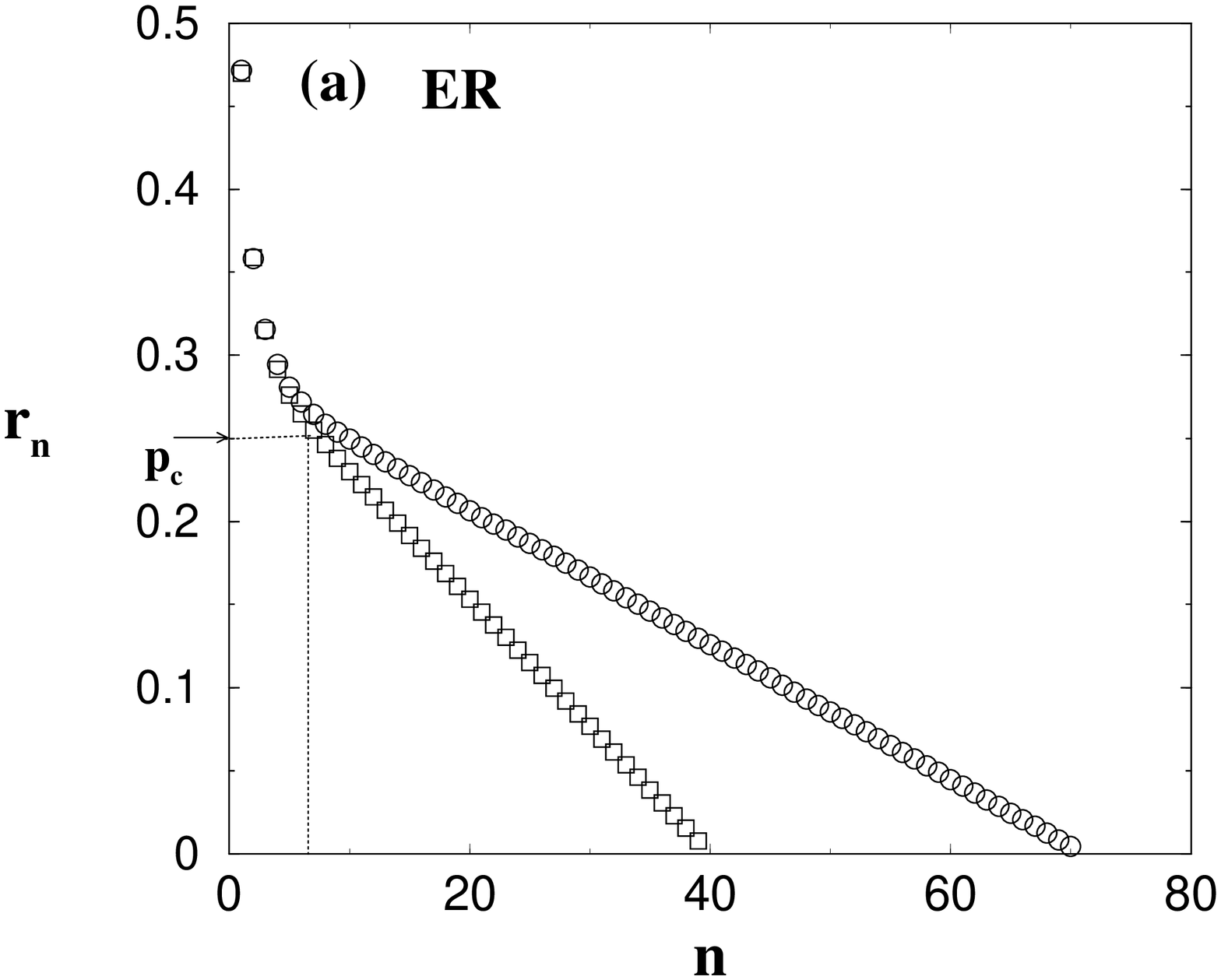}
\includegraphics[width=6.0cm,height=8.0cm,angle=270]{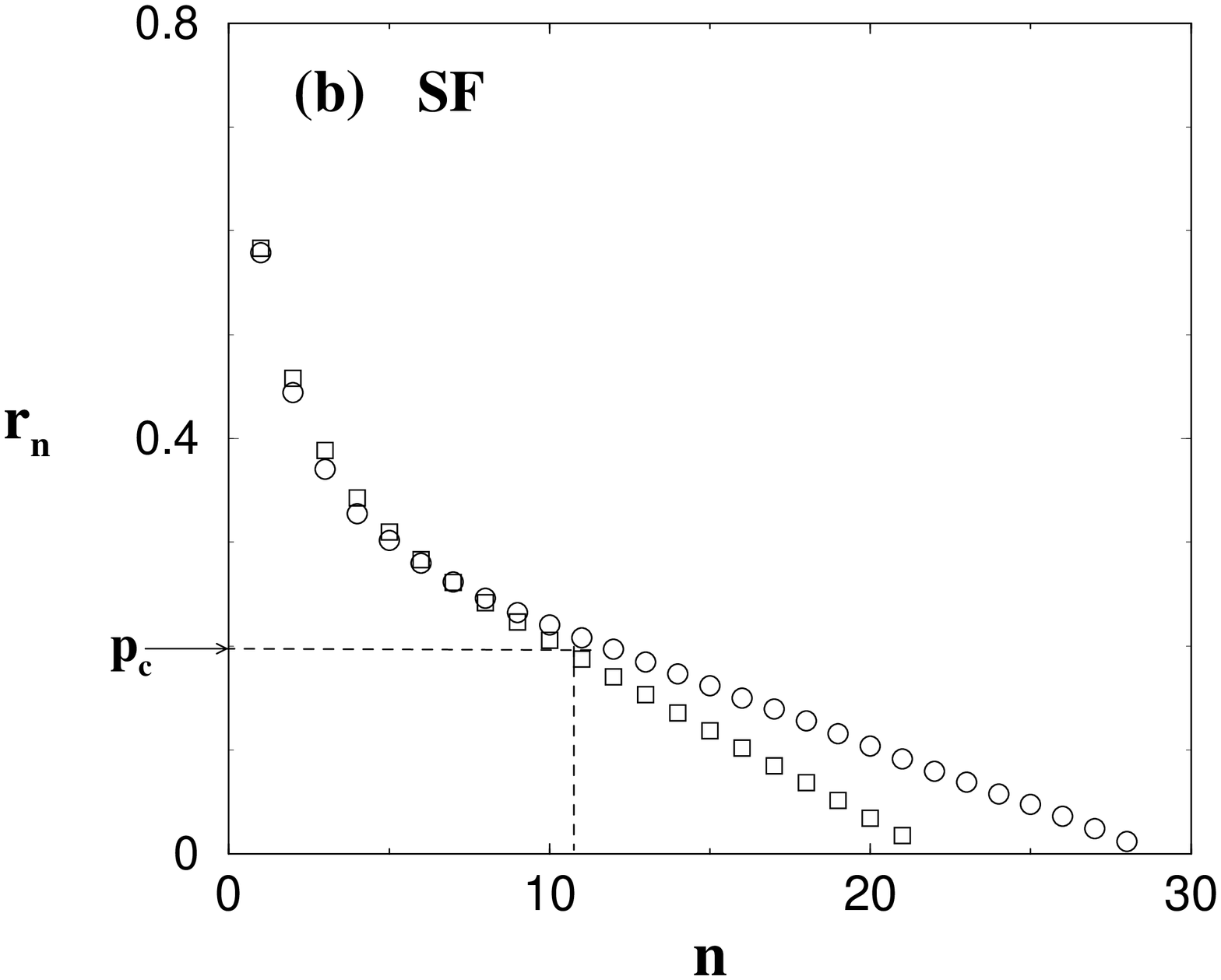}
\caption{Dependence on rank $n$ of the average values of the random
  numbers $r_n$ along the most probable optimal path for (a) ER random
  networks of two different sizes $N = 4096$ ($\Box$) and $N = 16384$
  ($\circ$) and, (b) SF random networks.}
\label{f.4}
\end{figure}

\begin{figure}
\includegraphics[width=6.0cm,height=8.0cm,angle=270]{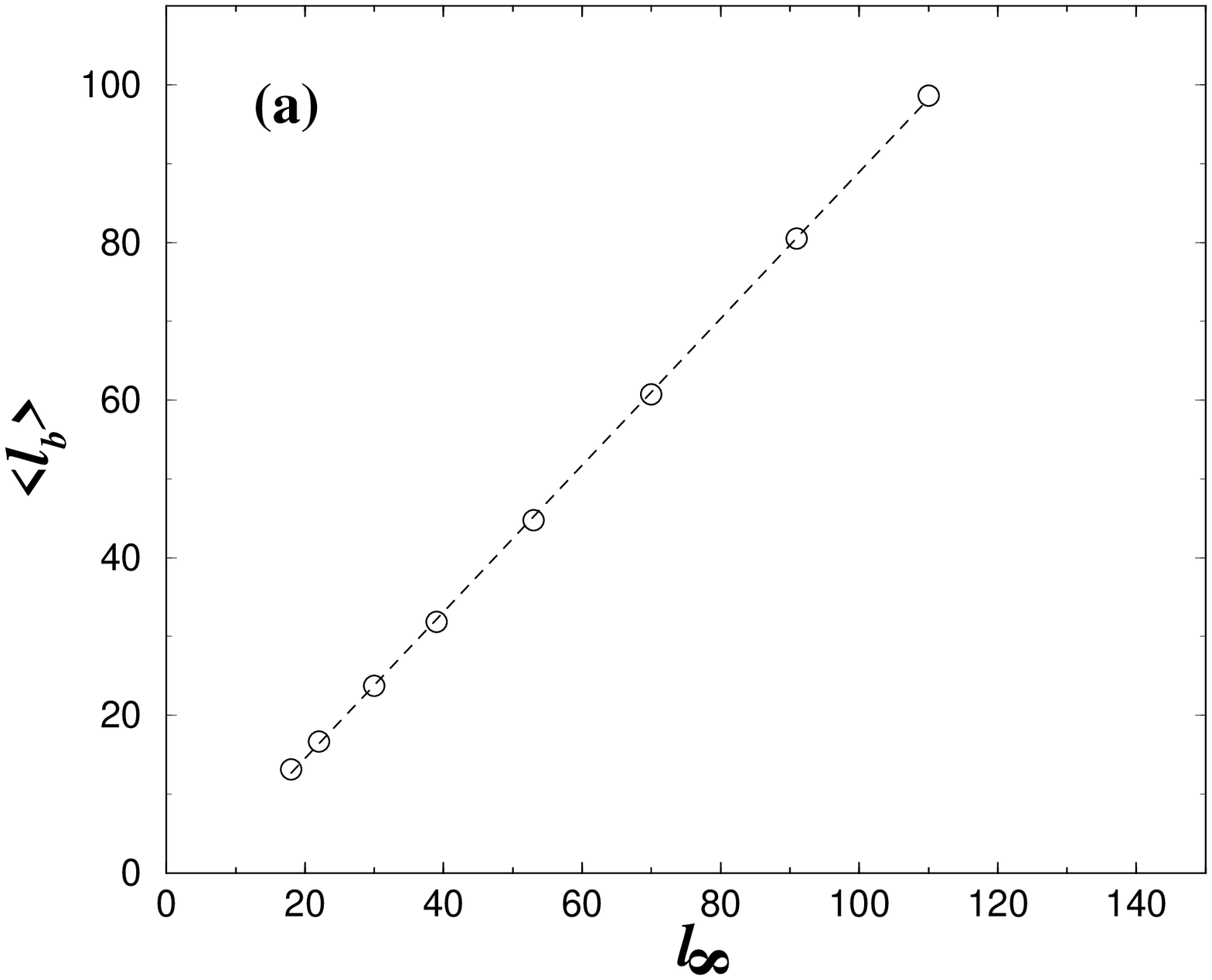}
\includegraphics[width=6.0cm,height=8.0cm,angle=270]{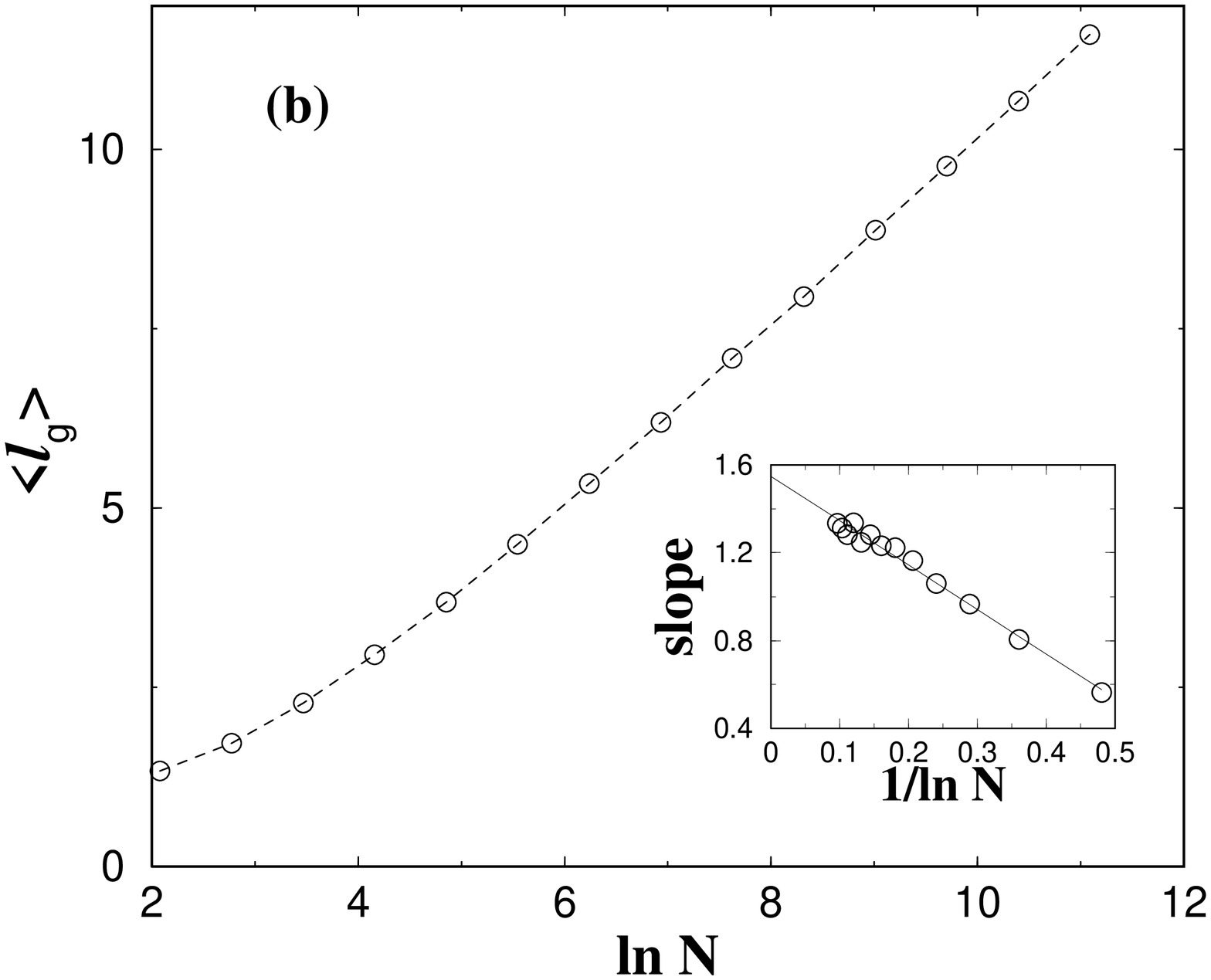}
\caption{(a) The average number of links $\langle\ell_b \rangle$ with
  random number values $r \le p_c$ on the min-max path plotted as a
  function of its length $\ell_\infty$ for an ER network, showing that
  $\langle\ell_b \rangle$ grows linearly with $\ell_\infty$. (b) The
  average number of links $\langle\ell_g\rangle$ with random number
  values $r>p_c$ on the min-max path versus $\ln N$ for an ER
  network, showing that $\langle\ell_g\rangle\sim\ln N$. The
  inset shows the successive slopes, indicating
  that in the asymptotic limit $\langle\ell_g\rangle \approx 1.55\ln N$.}
\label{f.5}
\end{figure}

We begin by making few observations about the min-max path. In
Fig.~\ref{f.4} we plot the average value of the random numbers $r_n$ on
the min-max path as a function of their rank $n$ ($1 \le n
\le\ell_{\infty}$) for ER networks with $\langle k \rangle = 4$ and for
SF networks with $\lambda = 3.5$. This can be done for a min-max path of
any length but in order to get good statistics we use the most probable
min-max path length. We call links with $r \le p_c$ ``black'' links, and
links with $r>p_c$ ``gray'' links, following the terminology of
Ioselevich and Lyubshin \cite{Ios04} where $p_c$ is the percolation
threshold of the network \cite{Cohen00}.
\noindent
We make the following observations regarding the min-max path:

\begin{itemize}

\item[{(i)}] For $r_n<p_c$, the values of $r_n$ decrease linearly with
     rank $n$, implying that the values of $r$ for black links are
     uniformly distributed between $0$ and $p_c$, consistent with the
     results of Ref.~\cite{Szabo03}. This is shown in Fig.~\ref{f.4}.

\item[{(ii)}] The average number of black links, $\langle\ell_b
     \rangle$, along the min-max path increases linearly with the
     average path length $\ell_\infty$. This is shown in Fig.~\ref{f.5}a.

\item[{(iii)}] The average number of gray links $\langle\ell_g \rangle$
     along the min-max path increases logarithmically with the average
     path length $\ell_\infty$ or, equivalently, with the network size
     $N$. This is shown in Fig.~\ref{f.5}b.

\end{itemize}
\noindent
The simulation results presented in Fig.~\ref{f.5} pertain to ER
networks; however, we have confirmed that the observations (ii) and (iii) also
hold for SF networks.
Next we will discuss our observations using the concept of the optimal
spanning tree. The {\it optimal spanning tree\/} (OST) is a subset of
links of a connected graph which provides an optimal path from node $A$
(which serves as the root of the tree) to any other node on the
graph. When the total weight of this path is dominated by the largest
weight of the links along the path (strong disorder limit), the OST does
not depend on the root and is determined only by the structure of the
original graph and a particular realization of the disorder. In this
limit, the OST becomes identical to the {\it minimal spanning tree\/}
(MST) \cite{Szabo03,Dob}. The path on the MST between any two nodes $A$
and $B$, is the optimal path between the nodes in the strong disorder
limit---i.e, the min-max path.

To construct the MST, we remove links in the descending order of their
costs $\tau_i$. If removal of a link destroys the connectivity of the
graph, we restore that link. This procedure is continued till there are
exactly $N-1$ links remaining. At this point the number of remaining
black links is
\begin{equation}
N_b={N\langle k\rangle p_c\over 2},
\label{equation6}
\end{equation}
where $\langle k \rangle$ is the average degree of the original graph
and $p_c$ is given by \cite{Cohen00}
\begin{equation}
p_c={\langle k \rangle\over\langle k^2 -k\rangle}.
\label{equation7}
\end{equation}

\begin{figure}
\includegraphics[width=8.0cm,height=5.0cm,angle=0]{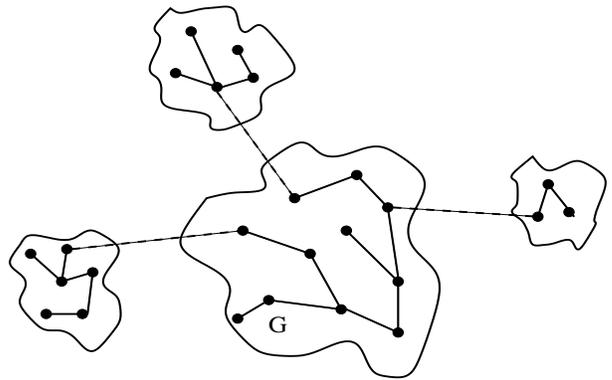}
\caption{Schematic representation of the structure of the minimal
  spanning tree, at the percolation threshold, with G being the giant
  component. Inside each cluster, the nodes are connected by black links
  to form a tree. The dotted lines represent the gray links which
  connect the finite clusters to form the gray tree. In this example
  $N_c = 4$ and the number of gray links equals $N_c - 1 = 3$.}
\label{f.6}
\end{figure}

The black links give rise to $N_c$ disconnected clusters. One of these
is a spanning cluster, called the {\it giant component}. The $N_c$
clusters are linked together into a connected tree by exactly $N_c-1$
gray links (see Fig.~\ref{f.6}). Each of the $N_c$ clusters is itself a
tree, since a random graph can be regarded as an infinite dimensional
system, and at the percolation threshold in an infinite dimensional
system the clusters can be regarded as trees. Thus the $N_c$ clusters
containing $N_b$ black links, together with $N_c-1$ gray links form a
spanning tree consisting of $N_b + N_c -1$ links.

Thus the MST provides a min-max path between any two points on the
graph. Since the MST connects $N$ nodes, the number of links on this
tree must be equal to $N-1$, so
\begin{equation}
N_b + N_c = N.
\label{equation8}
\end{equation}
From Eq.~(\ref{equation6}) and Eq.~(\ref{equation8}) it follows that
\begin{equation}
N_c=N\left(1-{\langle k\rangle p_c\over 2}\right).
\label{equation9}
\end{equation}
Therefore $N_c$ is proportional to $N$.

The path between any two nodes on the MST consists of $\ell_b$ black
links. Since the black links are the links that remain after removing
all links with $r>p_c$, the random number values $r$ on the black links
are uniformly distributed between $0$ and $p_c$ in agreement with
observation (i) and Ref.~\cite{Szabo03}.

Since there are $N_c$ clusters which include clusters of nodes connected
by black links as well as isolated nodes, the MST can be described as an
effective tree of $N_c$ nodes, each representing a cluster, and $N_c -
1$ gray links. We call this tree the ``gray tree'' (see
Fig.~\ref{f.6}). This tree is in fact a scale free tree \cite{Tomer04}
with degree exponent $\lambda_g = 2.5$ for ER networks and for scale for
networks with $\lambda \ge 4$, and  $\lambda_g =
(2\lambda-3)/(\lambda-2)$ for SF networks with $3 < \lambda < 4$
\cite{exp1}. If we take two nodes $A$ and $B$ on our original network,
they will most likely lie on two distinct effective nodes of the gray
tree. The number of gray links encountered on the min-max path
connecting these two nodes will therefore equal the number of links
separating the effective nodes on the gray tree. Hence the average
number of gray links $\langle\ell_g\rangle$ encountered on the min-max
path between an arbitrary pair of nodes on the network is simply the
average diameter of the gray tree. Our simulation results (see
Fig.~\ref{f.5}b) indicate that
\begin{equation}
\label{equation9x}
\langle\ell_g \rangle \sim \ln N.
\end{equation}
%
Since $\langle\ell_g\rangle\sim\ln\ell_\infty\ll\ell_\infty$, the
average number of black links $\langle\ell_b\rangle$ on the min-max path
scales as $\ell_\infty$ in the limit of large $\ell_\infty$ in agreement
with observation (2) as shown in Fig.~\ref{f.5}a. 

Now we will discuss the implications of our findings for the crossover
from strong to weak disorder. From observations (i) and (ii), it follows
that for the portion of the path belonging to the giant component, the
difference between consecutive random values $\Delta r$ is given by
$\Delta r=p_c/\langle\ell_b\rangle$. Hence we can approximate the sum of
weights of such links by the sum of the geometric progression
\begin{equation}
\sum^{\ell_b}_{k=1}\exp(ka\Delta r) =  {\exp(a\Delta
r)[\exp(ap_c)-1]\over\exp(a\Delta r)-1}\equiv\exp(ar^\ast) ,
\label{equation10}
\end{equation}
where $r^{\ast}\approx p_c+(1/a)\exp(-ap_c/\langle \ell_b \rangle)$
\cite{exp2}. Since $\langle\ell_b\rangle\approx\ell_\infty$,
\begin{equation}
r^{\ast}\approx p_c+{1\over a}\exp\left({-ap_c\over\ell_\infty}\right).
\label{equation11}
\end{equation} 
Thus restoring a short-cut link between two nodes on the optimal path
with $p_c <r< r^{\ast}$ may drastically reduce the length of the optimal
path. When $ap_c\gg\ell_{\infty}$ the probability that such a link
exists is negligible, but the probability becomes significant for
$ap_c\approx\ell_{\infty}$. Hence when the min-max path is of length
$\ell_{\infty} \approx ap_c$ the deviation of the optimal path from the
min-max path becomes significant. The length of the min-max path at
which the deviation occurs is precisely the crossover length $\ell^*(a)$,
and therefore $\ell^{\ast}(a) \sim ap_c$. 
Note that in the case of SF networks, as $\lambda
\to 3^+$, $p_c$ approaches zero and consequently $\ell^*(a)\to 0$. This
suggests that for any finite value of disorder strength $a$, a SF
network with $\lambda = 3$ is in the weak disorder regime.

In summary, for both ER random networks and SF networks we obtain a
scaling function for the crossover from weak disorder characteristics to
strong disorder characteristics. We show that the crossover occurs when
the min-max path reaches a crossover length $\ell^*(a)$ and $\ell^*(a)\sim
a$. Equivalently, the crossover occurs when the network size $N$ reaches
a crossover size $N^*(a)$, where $N^*(a)\sim a^3$ for ER networks and for SF
networks with $\lambda \ge 4$ and $N^*(a)\sim a^{\frac{\lambda -
1}{\lambda -3}}$ for SF networks with $3 < \lambda < 4$.

\subsubsection*{Acknowledgments}

We thank the Office of Naval Research, the Israel Science Foundation, and
the Israeli Center for Complexity Science for financial support, and
R. Cohen, E. Lopez, E. Perlsman, G. Paul, T. Tanizawa, and Z. Wu for
discussions.


\begin{thebibliography}{99}
\bibitem{exp1} This is a consequence of the fact that for the
original network the clusters at percolation have sizes $s$
distributed as $P(s) \sim s^{-\tau}$ \cite{Cohen02}, (with $\tau = 2.5$ for ER networks and for SF networks
with $\lambda \ge 4$, and  $\tau = (2\lambda-3)/(\lambda-2)$ for SF
networks with $3 < \lambda < 4$) and each node
within this cluster has a non-zero probability of connecting to a node
outside the cluster.
\end{thebibliography}
\end{document}